%% file: main.tex
\documentclass{article} 
\usepackage{graphicx}
\usepackage{amsmath}
\usepackage{booktabs}
\usepackage{authblk}
\usepackage{mathptmx}
\usepackage{color,soul}
\usepackage{subfig}
\usepackage{xcolor}
\usepackage[normalem]{ulem}

\usepackage{hyperref}
\hypersetup{
    colorlinks=true,
    urlcolor=blue,
    citecolor=blue,
    linkcolor=blue,
}

\usepackage{url}
\makeatletter
\def\url@leostyle{%
  \@ifundefined{selectfont}{\def\UrlFont{\sf}}{\def\UrlFont{\small\sffamily}}}
\makeatother
\usepackage[
  backend=biber,
  maxbibnames=9,  
  maxcitenames=2, 
  uniquename=false, 
  uniquelist=false, 
  style=numeric-comp, 
  sortcites=true, 
]{biblatex}
\date{}

\title{Reservoir Computing with Magnetic Thin Films}

\author[1]{Matthew Dale}
\author[1]{David Griffin}
\author[2]{Richard F. L. Evans}
\author[2]{Sarah Jenkins}
\author[1]{Simon O'Keefe}
\author[3]{Angelika Sebald}
\author[1,*]{Susan Stepney}
\author[1]{Fernando Torre}
\author[2]{Martin A. Trefzer}

\affil[1]{Department of Computer Science, University of York, UK}
\affil[2]{School of Physics, Engineering and Technology, University of York, UK}
\affil[3]{Department of Chemistry, University of York, UK}
\affil[*]{susan.stepney@york.ac.uk}

\begin{document}

\maketitle



\begin{abstract} 

Advances in artificial intelligence are driven by technologies inspired by the brain, but these technologies are orders of magnitude less powerful and energy efficient than biological systems. Inspired by the nonlinear dynamics of neural networks, new unconventional computing hardware has emerged with the potential 
to exploit natural phenomena and gain efficiency, in a similar manner to biological systems. Physical reservoir computing demonstrates this with a variety of unconventional systems, from optical-based to memristive systems. Reservoir computers provide a nonlinear projection of the task input into a high-dimensional feature space by exploiting the system's internal dynamics. A trained readout layer then combines features to perform tasks, such as pattern recognition and time-series analysis. Despite progress, achieving state-of-the-art performance without external signal processing to the reservoir remains challenging. Here we perform  an initial exploration of three magnetic materials in thin-film geometries via micro-scale simulation. Our results reveal that basic spin properties of magnetic films generate the required nonlinear dynamics and memory to solve machine learning tasks (although there would be  practical challenges in exploiting these particular materials in physical implementations). 
The method of exploration can be applied to other materials, so this work opens up the possibility of testing different materials, from relatively simple (alloys) to significantly complex (antiferromagnetic reservoirs). 
\end{abstract} 

keywords: reservoir computing; magnetic computing; edge computing

\input{main_content}

\section*{Acknowledgements}
This work was carried out under the SpInspired project, EPSRC Grant EP/R032823/1,
and the MARCH project, EPSRC grant : EP/V006029/1. All experiments were carried out using the Viking Cluster, a high performance compute facility provided by the University of York. FT acknowledges funding from the 2019 YCCSA (York Cross-disciplinary
Centre for Systems Analysis) summer school.






\printbibliography

\appendix
\input{supplementary}

\end{document}

%% file: main_content.tex
\section{Introduction} \label{sec:Introduction}

Performing machine learning at `the edge' is a growing area of interest, where inference is performed locally in real time ~\cite{shi2016edge,chen2019deep,wang2020convergence}. Embedded devices that can perform complex information processing without the need to outsource to remote servers are ideal for real-time applications. Current digital technology is limited by processing speeds, memory, size, and power consumption. Unconventional hardware is a potential alternative to classical computing hardware, with low-energy consumption, inherent parallelism, and no separation between processor and memory (the von Neumann bottleneck)~\cite{adamatzky2016advancesv2}. Neuro-inspired hardware~\cite{young2019review} is one possible route to embed machine learning at the edge, another is to exploit embodied computation in novel dynamical systems.

By design, neuromorphic hardware implements the abstract behaviour of neurons and their connectivity at the lowest circuit level, e.g.\ weighted summation, threshold functions, synapses. This typically requires a combination of simpler components to implement the model. For example, a single neuron with conventional complementary metal–oxide–semiconductor technology takes 10s to 100s of transistors to replicate a neuron-synapse circuit~\cite{indiveri2011neuromorphic,jo2010nanoscale}. Another option is to force the neuron model directly onto the material to improve energy-efficiency and reduce the physical footprint, for example, using memristor technology~\cite{caravelli2018memristors,xia2019memristive,Ye2022}. However, neural-like representations (e.g. spike trains) and model constraints may require removal of useful natural properties (e.g.\ variability in components) or require additional engineering such as material optimisation or adding artificial memory.

Here we take an alternative approach, exploiting the dynamical behaviours of nonlinear systems. As dynamical properties, such as memory and nonlinear oscillation, can occur naturally in complex materials, using these properties directly promises to be a more efficient approach than implementing specific neural units.

The discovery and control of these material properties, some of which may be intractable or unknown, raises new challenges. Two novel approaches have been proposed to exploit the embodied computation of materials for computing: evolution \textit{in materio} and reservoir computing. 

Evolution \textit{in materio} uses computer-controlled manipulation of external stimuli to configure the material and its input-output mapping, using digital computers to directly evolve physical material configurations. Miller and Downing~\cite{miller2002evolution} proposed using artificial evolution as a mechanism to exploit and configure materials, arguing natural evolution is the method \textit{par excellence} for exploiting the physical properties of materials. A range of materials have been evolved to perform various computational tasks, such as classification, real-time robot control and pattern recognition~\cite{mohid2015EvoJournal,massey2016SciRep,bose2015evolution,chen2020classification}. 

Reservoir computing is a neuro-inspired framework that harnesses the high-dim\-ension\-ality and temporal properties of random recurrent networks and novel systems \cite{schrauwen2007overview,verstraeten2009quantification}.  Reservoir computers provide a nonlinear projection of the task input into a high-dimensional spatial-temporal feature space by exploiting the system's internal dynamics. A trained readout layer then combines features to perform tasks, such as pattern recognition and time-series analysis. Physical implementations of the reservoir model are diverse~\cite{appeltant2011information, dion2018reservoir,moon2019temporal}. Recent spintronic reservoirs show some key advantages compared to other systems, combining GHz+ operating frequencies, ultra-compact size and ultra-low-energy consumption \cite{Allwood2023,prychynenko2018magnetic,pinna2020reservoir,torrejon2017neuromorphic,nakane2018reservoir,romera2018vowel,zheng2020recurrent,watt2020reservoir,zahedinejad2020two}.

Here we explore, in simulation, material computation by ferromagnetic materials in thin nano-film geometries, combining both evolution \textit{in materio} and reservoir computing methods. The reservoir model is used to harness the propagation of information through magnetic films, and artificial evolution is used to optimise the material reservoir. 
Using open-source micro-scale simulation software, we evolve properties of three ferromagnetic materials to solve three time-dependent tasks of increasing complexity. 
The exploration finds optimal parameters for these tasks, and is compared to relevant state of the art neural networks of comparable size. The magnetic system is then characterised by task-independent measures to distinguish and understand the dynamical properties of each material. 
As the initial exploration is conducted without the effects of thermal noise, a further set of experiments is conducted to characterise the effects of thermal noise and its interaction with the film size.
Lastly, we discuss the process of experimentally evaluating the computing capabilities of any new system and potential caveats. We argue that task-independent characterisation should ideally be performed early in any design cycle to determine appropriate benchmarks, parameters, and future applications.

\begin{figure}[tp]
  \centering
\includegraphics[width=0.95\linewidth,trim=0.25cm 0.25cm 0.25cm 0.5cm,clip]{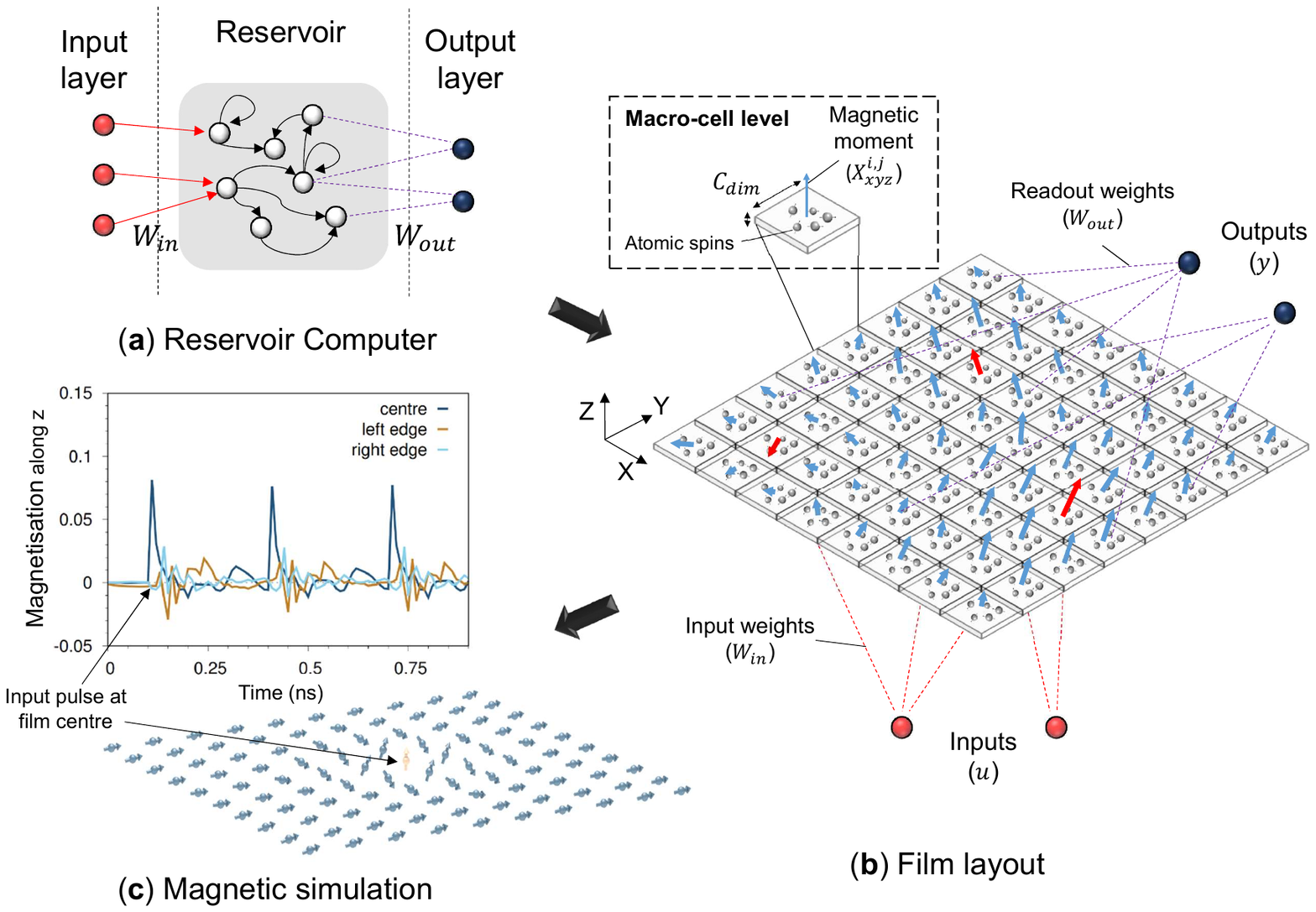}
\caption{
(a) Reservoir computing model split into input, reservoir and output layers connected by adjustable weights. The reservoir is self-contained, typically featuring a sparse, recurrent network of processing nodes. 
(b) Schematic of simulated thin-film magnetic reservoir system, consisting of micromagnetic macro-cells with properties derived from atomistic values. Global input sources $u$ connect via weights $W_{in}$ to drive local magnetisation fields inducing spin oscillations. Each macro-cell's average magnetic moment produces a 3-d orientation vector $X_{xyz}$ forming a reservoir state. States are then combined via a linear readout function $W_{out}$ to produce the final system output $y$. 
(c) Impulse response of micromagnetic spin system. Signal injected in the centre of the film via the $z$-axis at 25 time-step intervals with 10\,ps scanning frequency. 
}
\label{fig: magnetic platform}
\end{figure}

\section{Reservoir Computing System}

Reservoir computers are composed of three layers: input, reservoir and output layer (Fig.~\ref{fig: magnetic platform}a). A reservoir, typically a fixed random network of discrete processing nodes with recurrent connections, features non-linear characteristics and a short-term memory. The reservoir network is driven by a time-varying input \textbf{$u$} that propagates through an input mapping given by typically random connection weights {$\mathbf{W}_{in}$}. The non-linear reservoir provides a high-dimensional projection of the input, from which a subsequent linear readout layer can extract features relevant to the problem task. 

The reservoir dynamics of  leaky-integrated discrete-time continuous-value simulated networks are commonly given by the state update equation:
\begin{eqnarray} \label{eq: esn state update}
\hat{\mathbf{x}}(n) &=& f(b\mathbf{W}_{in}[u(n); u_{bias}] + c\mathbf{W}\mathbf{x}(n-1))
\\  \label{eq: esn state update2}
\mathbf{x}(n) &=& (1-a)\mathbf{x}(n-1) + a\hat{\mathbf{x}}(n) 
\end{eqnarray}
where $\mathbf{x}$ is the internal state at sample time $n$ ($\hat{\mathbf{x}}$ is an auxiliary variable capturing the updated state ignoring leakage).
The non-linear activation function $f$ is typically a \textit{tanh} function.
The input signal \(u\) is augmented with  a bias source $u_{bias}$.
\(\mathbf{W}_{in}\) and \(\mathbf{W}\) are typically random weight matrices giving the connection weights to inputs and internal neurons respectively. 
Scaling parameters $b$ and $c$ control the global scaling of the input weights and reservoir weights respectively. Input scaling $b$ affects the non-linear response of the reservoir and relative effect of the current input. Internal scaling $c$ controls the reservoir's stability as well as the influence and persistence of the input: low values dampen internal activity and increase response to input, and high values lead to chaotic behaviour. 
The leakage filter $a$ is used to match the internal timescales of the film to the characteristic timescale of the task. This is similar to adding a low-pass filter before the output. The leak rate controls the timescale mismatch between the input and reservoir dynamics; when $a=1$, the previous states do not leak into the current states.  

The trained output \(y(n)\) is given when the reservoir states \(\mathbf{x}(n)\) are viewed through the trained readout weight matrix \(\mathbf{W}_{out}\): 
\begin{equation} \label{eq: esn output}
y(n)= \mathbf{W}_{out}x(n)
\end{equation}

Training occurs only at the readout stage,
through trained weighted connections $\mathbf{W}_{out}$ connecting observable states to the final output. Typically, one-shot learning is used through linear regression, making learning fast.

Training attempts to minimise the error \(E(y, y^{target})\) between the reservoir output $y$ and the target output $y^{target}$. 
Here we use the Normalised Mean Square Error (NMSE) measure:
\begin{equation} \label{eq: nmse}
NMSE = \frac{\langle (y^{target}-y)^2\rangle}{Var(y^{target})}
\end{equation}
where $\langle.\rangle$ denotes the mean over multiple time steps, and the mean square error is normalised with the variance $Var(.)$ of the target output.

\section{Thin Film Reservoir Computer}

Fig.~\ref{fig: magnetic platform}b shows the layout of the simulated magnetic system and its reservoir representation. The film is conceptually divided into a grid of \textit{macro-cells} (comprising the average properties of multiple atoms) for the purpose of simulation and to define input-output locations; the underlying film  does not possess discrete processing nodes.
The macro-cells in all simulations have an area of 5\,nm $\times$ 5\,nm;
for the initial experiments, macro-cells are  one unit cell thick (see table~\ref{tab:material param});  thicker films are explored in section~\ref{sec:paths}. 

Each macro-cell is connected to a time-varying input signal source and a bias source via weighted connections $\mathbf{W}_{in}$. The physical input is encoded as a magnetic field that interacts with the film induced by an electrical current. Inputs can be physically implemented in various ways, for example, using a single source connected to individual amplifiers supplying localised magnetic fields. The output of each macro-cell is represented by a three-dimensional magnetisation vector $X_{xyz}$. This approach models a grid of nano-contacts across the film, measuring a low-resolution snapshot of the film's magnetic state. 

The material is interpreted as a reservoir in the following way:
\begin{eqnarray} \label{eq: material eq}
\mathbf{X} &=& \sigma(b\mathbf{W}_{in}[u; 1], \alpha)
\\
 \mathbf{X}_f(n) &=& (1-a)\mathbf{X}(n-1) + a \mathbf{X}(n) 
\\
y(n) &=& \mathbf{W}_{out}\mathbf{X}_f(n)
\end{eqnarray}
where $\mathbf{X}$ is the global material state, with components formed from each macro-cell's local $X_{xyz}$ 3d magnetisation vector.
$\sigma$ represents the non-linear magnetic material function, a property of the specific material.
The input scaling parameter $b$ applied to the input mapping is captured by the field intensity ($0<b\leq2$). This suppresses or raises the overall magnitude of the locally applied fields, promoting varying dynamical behaviours. 
The input mapping defined by \(\mathbf{W}_{in}\) consists of weighted connections to each macro-cell from the input vector $u$ with a bias of 1. 
In a physical device this mapping could be calculated in software prior to application to the device,
or achieved physically using suitable analogue circuitry. 
The magnetic damping parameter ($0<\alpha\leq1$) controls the speed of information propagation and oscillation. Damping describes the non-linear spin relaxation across the film, controlling the rate at which magnetisation spins reach equilibrium. 

$\mathbf{X}_f$ is an external filter layer with a one-step memory implemented after the observation of material state $\mathbf{X}$ and before the readout weights $\mathbf{W}_{out}$ are applied; 
$a$ is the leakage parameter.
As the leak filter is provided post state observation, it does not affect the internal dynamics of the system, unlike in Eq.~\ref{eq: esn state update}.
However, it served the same purpose: an additional tuning parameter to match timescales~\cite{dale2016reservoir,dale2017reservoir}.


Here we focus on three ferromagnetic metals: 
cobalt (Co), nickel (Ni) and iron (Fe). 
The atomic magnetic properties of these materials are well understood from first principle calculations~\cite{pajda2001ab}, providing a detailed insight into microscopic and macroscopic magnetic behaviour. 
The properties of these metals are well studied, so they provide a useful point of reference that prevents the evolutionary search from exploring completely unrealistic  parameter values. 

As a thin film, the reservoir is highly structured. The influence each simulated macro-cell has on its nearest neighbours is determined by the physical properties of exchange, anisotropy, and dipole Hamiltonian (see section~\ref{sec:spin model}). The exchange interactions dominate over short lengthscales, meaning that macro-cells have short range time- and spatial correlations over the total sample size. Fig.~\ref{fig: magnetic platform}c shows a typical simulated micromagnetic response to three input pulses at the film's centre. When perturbed, spin waves propagate through the film, inducing reflections, oscillations and interference patterns. At the edges, a similar characteristic response is seen per impulse, but with some contributions from previous stimuli.




\begin{figure}[tp]
\centering
\subfloat[$7\times7 = 49$ macro-cells]{\includegraphics[width=0.6\linewidth,height=0.5\linewidth,trim=3cm 1.5cm 3cm 2.75cm,clip]{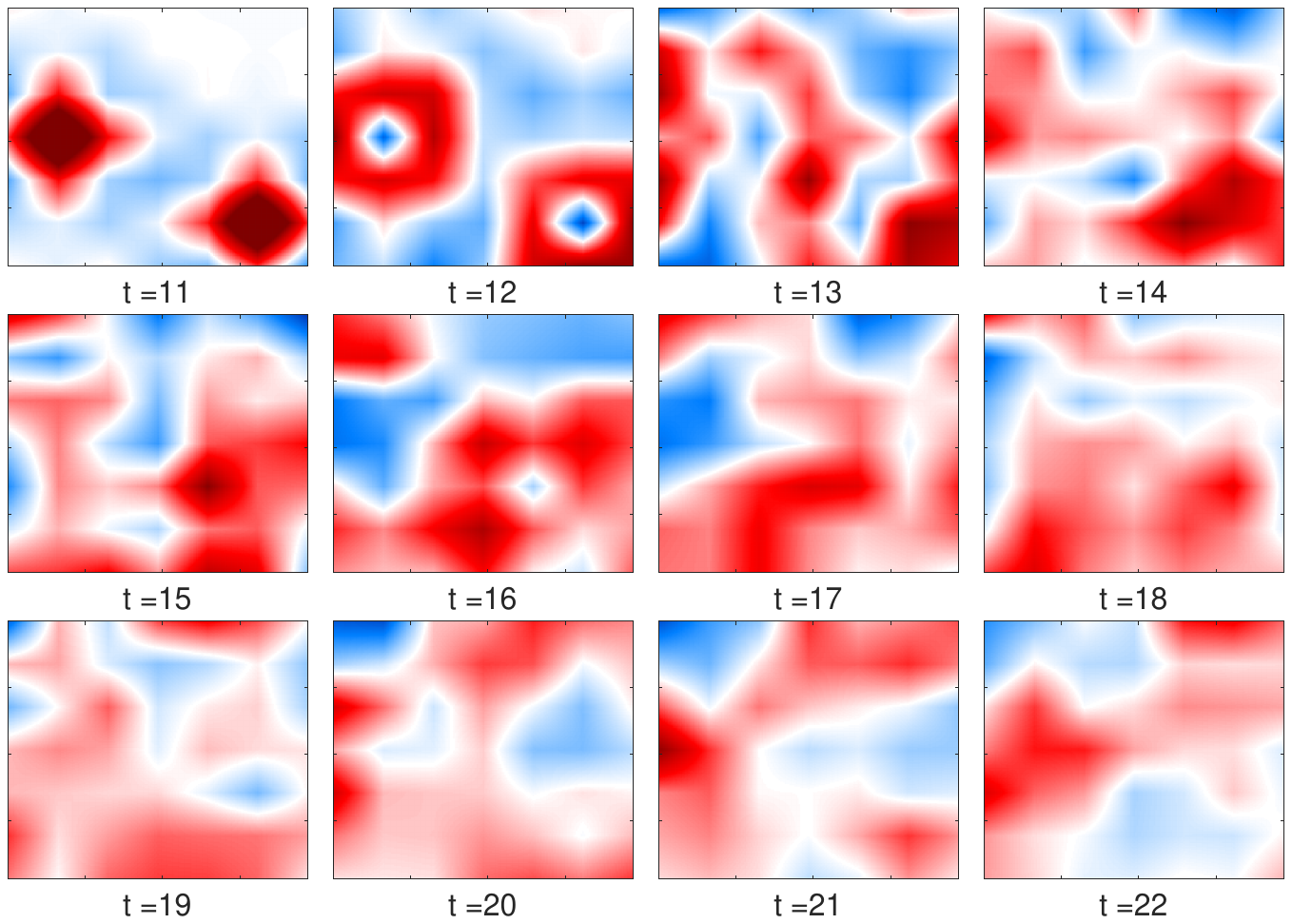}\label{fig: 49 cell input pulse}} \hspace{0.5cm}
\subfloat[$30\times 30 = 900$ macro-cells]{\includegraphics[width=0.6\linewidth,height=0.5\linewidth,trim=3cm 1.5cm 3cm 2.75cm,clip]{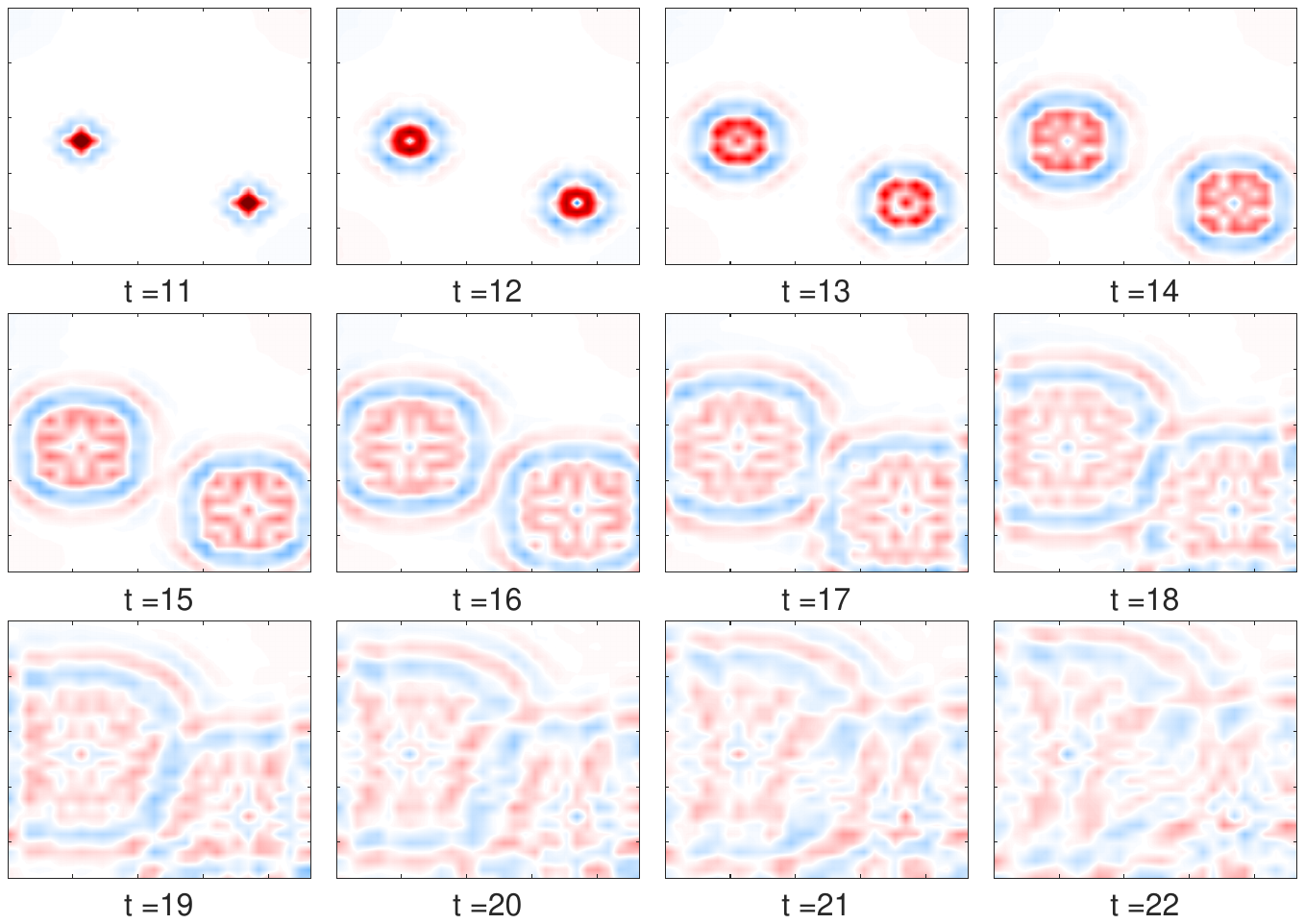}\label{fig: 900 cell input pulse}}
\caption{Dynamics of Co magnetic film at different sizes. An input pulse is supplied at two locations on the film at $t=10$. Red indicates a positive magnetisation, and blue, negative.}
\end{figure}

As information propagates through the film, local coupled spins form wave crests and troughs that interact, creating interference patterns. At the boundaries, waves are reflected back into the film. Figs.~\ref{fig: 49 cell input pulse} and~\ref{fig: 900 cell input pulse} provide a visualisation of this dynamical behaviour for two Co films (49 macro-cell and 900 macro-cell). At $t=10$, a single input pulse is supplied to two separate input locations. At $t=11$--$13$, waves appear and propagate. The smaller film (Fig.~\ref{fig: 49 cell input pulse}) interacts almost instantly with the boundaries, and waves reverberate around the film for some time. In the larger film (Fig.~\ref{fig: 900 cell input pulse}), signals propagate for longer, undisturbed, until the wave crests reach each other and the boundaries. At $t>20$, interference and reflected waves begin to dominate; however, memory of past inputs are still recoverable. 

\subsection{Magnetic Model} \label{sec:spin model} 

The reservoir thin films are simulated micromagnetically where the atomistic detail is coarse grained into macro-cells. For a generic atomistic model with $n$ nearest neighbour interactions, the Curie temperature $T_C$ can be calculated from the atomistic exchange $J_{ij}$ by the mean-field expression. This sums over every exchange that occurs in each macro-cell to calculate the total exchange~\cite{jiles2015introduction}.

\begin{equation}
T_C = \frac{\varepsilon}{3 k_{\mathrm{B}} N_{c}}\sum_{i = 0}^{N_{c}} \sum_{j =0}^n J_{ij}
\end{equation}
where $k_B$ is the Boltzmann constant, $N_{c}$ is the number of atoms per macro-cell, and $\varepsilon$ is a correction factor from the usual mean-field expression which arises due to spin waves in the 3D Heisenberg model. 

The anisotropy $k_u$ and the spontaneous magnetisation $M_s$ are calculated as a sum of the atomic anisotropies and spin moments within each macro-cell. The gyromagnetic ratio $\gamma$ and the damping constant $\alpha$ are calculated as an average of the atomic parameters for each macro-cell. 

The energetics of the micromagnetic system are described using a spin Hamiltonian neglecting non-magnetic contributions and given by:
\begin{equation}
\centering
    \mathcal{H}_{eff} = \mathcal{H}_{app} + \mathcal{H}_{ani} + \mathcal{H}_{exc} +\mathcal{H}_{dip}
\end{equation}
where $\mathcal{H}_{app}$ is the applied field, $\mathcal{H}_{ani}$ is the anisotropy field, $\mathcal{H}_{exc}$ is the intergranular exchange, and $\mathcal{H}_{dip}$ is the dipole field. 

The anisotropy Hamiltonian describes the directional dependence of the material's magnetisation; in this case the anisotropy is uniaxial along $z$ and is described by:
\begin{equation}
\mathcal{H}_{ani} =  KV(\mathbf{m}_x^2 + \mathbf{m}_y^2)
\end{equation}

The exchange field is calculated as a sum of the exchange interactions between neighbouring macro-cells, the micromagnetic exchange constant $A$ is a sum over all atoms which have neighbours in another macro-cell. The summation over all the interactions gives a total interaction from macro-cell $i$ to macro-cell $j$. From this the micromagnetic exchange constant is calculated by multiplying by the distance between the atomistic atoms.

\begin{equation}
\mathbf{H}_{ex}^i = A_{ij} \frac{2}{M_S\, \Delta^2\, m_e^2} \sum_{n_{cells}} (\mathbf{m}_j - \mathbf{m}_i)
\end{equation}

\begin{equation}
\mathbf{H}_{dip} = \frac{\mu_0}{4\pi}   \frac{3 (\mathbf{m} \cdot \hat{\mathbf{r}}) \hat{\mathbf{r}} - \mathbf{m}}{|\hat{\mathbf{r}}|^3} - \frac{\mu_0 \mathbf{m}}{3V}
\end{equation}

The atomistic Landau–Lifshitz–Gilbert (LLG) equation is used to model the time-dependent behaviour of the magnetic films given by:
\begin{equation}
\frac{\partial \mathbf{m}_{i}}{\partial t}=-\frac{\gamma}{\left(1+\lambda^{2}\right)}\left[\mathbf{m}_{i} \times \mathbf{H}_{\mathrm{eff}}^{i}+\lambda \mathbf{m}_{i} \times\left(\mathbf{m}_{i} \times \mathbf{H}_{\mathrm{eff}}^{i}\right)\right]
\label{eq:micromagneticllg}
\end{equation}
where $\mathbf{m}_i$ is a unit vector representing the direction of the magnetic spin moment of macro-cell $i$, $\gamma$ is the gyromagnetic ratio and ${\mathbf {H}}^i_{{\rm eff}}$ is the net magnetic field on each macro-cell and is equal to the derivative of the spin Hamiltonian:
\begin{equation}
    \mathbf{H}_{\mathrm{eff}}^{i}=-\frac{1}{\mu_{\mathrm{s}}} \frac{\partial \mathcal{H}_{eff}}{\partial \mathbf{S}_{i}}
\end{equation}

The discrete micromagnetic macro-cells of the spin model represent reservoir ``nodes'' in our model, however traditional properties and notions such as node coupling and node dynamics (e.g. nonlinearity) do not commute. Instead, analogous mechanisms are explained by the spin Hamiltonian, for example, node state is simply the average local magnetic state of the film and node connectivity is defined by the material's exchange constant and dipole fields.

\section{Experimental Procedure and Parameters} \label{sec:setup}

\subsection{Search algorithm}

The  goal is to find parameter values that optimise system performance as a reservoir.
Many meta-heuristic search algorithms can be used to optimise reservoirs~\cite{bala2018applications}.
Here we use the microbial genetic algorithm (MGA)~\cite{harvey2009microbial}, 
due to its simplicity of implementation. The MGA allows individuals to survive across many generations, provides elitism for free, and offers a simple mechanism for selection, recombination and mutation. 

Parameters used in our MGA runs are: population size $=100$; number of generations\footnote{
Note that in the MGA, a `generation' corresponds to the production of a single new individual,
as opposed to more traditional evolutionary algorithms (EAs), where a generation corresponds to the production of an entire population of new individuals.
Hence 2000 MGA generations corresponds to approximately 20 traditional EA generations with the same population size of 100.
} $=2000$; 
mutation rate $=0.05  (5\%)$; recombination rate $=0.5 (50\%)$; deme size (species separation) $=0.1 (10\%$ of population); number of runs $=20$. 
These parameters are used for all experiments involving an evolutionary algorithm.
The result of an experiment is a set of 20 parameter values,
from the best-performing reservoir of each run.

We also use a random search algorithm for some comparison experiments.
To provide a fair comparison with the evolved solutions,
one run of random search examines 2000 randomly generated individuals.

Readout training is performed using ridge regression~\cite{lukovsevivcius2012practical} and occurs within the evolutionary loop during the training phase. A validation and testing phase is carried out to evaluate the generalisation of the readout to new data. This approach is similar to previous work~\cite{dale2016evolving,dale2018neuroevolution}.

\subsection{Material simulations}

During simulation, material parameter values such as exchange interaction, an\-iso\-tropies, and atomic moments are defined by the material and remain unaltered. The simulation parameters are given in Table~\ref{tab:material param}. These include exchange constants and second-order uniaxial anisotropy constants. To conduct accurate temperature calculations (section~\ref{sec:paths}), rescaling exponents and Curie temperature information are also included.

\begin{table}[tp]
\centering
\resizebox{1\columnwidth}{!}{%
\begin{tabular}{lllll}
\toprule
  & Ni & Co & Fe & unit\\
\midrule
crystal structure & fcc & fcc & bcc & --\\
unit cell size $a$ & 0.3524 & 0.2507 & 0.2866 & nm\\
atomic spin moment $\mu_s$ & 0.606 & 1.72 & 2.22 & $\mu_B$ \\
exchange energy $J_{ij}$ & $2.757\times 10^{-21}$ & $6.064\times 10^{-21}$ & $7.050\times 10^{-21}$ & J/link\\
anisotropy $k$ & $5.47\times 10^{-26}$ & $6.69\times 10^{-24}$ & $5.65\times 10^{-25}$ & J/atom\\
Temp. rescaling exponent  & $2.322$ & $2.369$ & $2.876$ & --\\
rescaling Curie temperature  & $635$ & $1395$ & $1049$ & --\\
\bottomrule
\end{tabular}%
}
\caption{\label{tab:material param} Physical parameters used in the simulation of each ferromagnetic material in VAMPIRE. These parameter values are static, and are not affected by the evolutionary algorithm.}
\end{table}

In the following magnetic material experiments the evolutionary algorithm tunes the following parameter values: the input mapping $W_{in}$, field intensity input scaling $b$, intrinsic magnetic damping $\alpha$, and post-state collection filter leakage $a$. Details on the initial values for these parameters are given in Table \ref{tab:initial-values}.

\begin{table}
    \centering
    \begin{tabular}{cll}
         \hline
         parameter  & description & initial value \\
         \hline
         $\alpha$ & intrinsic magnetic damping & uniform, $0.001$ and $1$ \\
         $a$ & post-state collection filter leakage & uniform, $0$ and $1$ \\
         $b$ & field intensity input scaling & uniform, $-2$ and $2$\\
         \hline
    \end{tabular}
    \caption{Initial parameter values used in the evolutionary algorithm.}
    \label{tab:initial-values}
\end{table}

To conduct the experiments, the VAMPIRE source code was augmented with a dynamic input-output mechanism. Important parameters for the VAMPIRE simulation include input frequency, integration time-step, initial spin direction, and macro-cell size (micromagnetic simulation). 

To exploit the fast spin dynamics of the ferromagnetic materials, data inputs are applied at 10\,ps simulated time intervals (100\,GHz). Selecting a suitable input timescale depends on the material's dynamics. An input faster or slower than the system's intrinsic timescale alters the temporal dynamics and thus can affect settling times, refractory periods and memory in the system. The chosen input frequency was based on qualitative experiments in search of characteristic behaviours, such as fast response and a short settling time. Providing input at a rate of 100\,GHz would currently require a state-of-the-art implementation; in practice, either different materials or a different understanding of the dynamic properties of the material may be necessary. 

Ideally, small integration timesteps are preferable to more accurately capture spin precession and general dynamics between input pulses; however, this comes with a large computational cost. 
To optimise the evaluation process and reduce computational cost an integration,
we use an integration timestep of 100\,fs. 
In preliminary experiments on material reservoirs with randomly set parameter values, we found that performance is not significantly affected by this choice (see Appendix~\ref{app: timestep}).

In material simulations, the initial spin direction is aligned with the $x$-axis, and input signals are injected in the $z$-direction adding to the current magnetic state of the film. Inputs through local magnetic fields interact with neighbouring macro-cells and dissipate depending on the material's spin coupling properties: increasing the magnetisation of a macro-cell increases its neighbours proportional to its distance and exchange properties.


\subsection{Comparison networks}
We compare the performance of magnetic film reservoirs with simulated recurrent neural networks of comparable size, where one macro-cell is assumed to correspond to one neural node.
The simulated networks use the state update defined in Eqs.~\ref{eq: esn state update},\ref{eq: esn state update2}. 

We use two comparison network connection topologies: a standard ESN (randomly connected), and a grid-connected network.
For the grid network, we define a square grid of nodes, each connected to its nearest neighbours in its Moore neighbourhood~\cite{adamatzky2010game}. Each non-perimeter node has eight connections to neighbours and one self-connection, resulting in each node having a maximum of nine adaptable weights in \(\mathbf{W}\).
Given our identification of a macro-cell with a network node, such a grid network more closely corresponds to the material  structure. 
Grid networks with recurrent connections have been shown to be dynamically similar to recurrent neural networks with less restrictive connectivity, but often have to compensate with larger network size~\cite{dale2019role,Dale++:2020:NACO,rodan2010simple}. 

We use two search algorithms:
random search and evolutionary search.
For both random and evolved reservoir networks, \(\mathbf{W}_{in}\) and \(\mathbf{W}\) are initialised as sparse matrices with input sparsity $0.1$, internal sparsity $0.1$, and values drawn from a normal distribution with mean $0$, variance $1$.



\subsection{Benchmark Tasks} \label{sec:BT} 
To evaluate the materials, three temporal tasks are applied. 
The chosen tasks are widely used benchmarks for different reservoir systems and methods~\cite{jaeger2001echo,rodan2010simple, tran2017memcapacitive, paquot2012optoelectronic, larger2012photonic, inubushi2017reservoir}.
Each benchmark increases in difficulty, demonstrating the film's dynamic range and ability to perform increasingly complex tasks.

The first benchmark is the time series prediction of a chaotic laser data set, chosen for its nonlinear properties and periodic structure.
The laser task predicts the next value of the Santa Fe time-series Competition Data (dataset A)~\cite{LaserData}. The dataset holds original source data recorded from a Far-Infrared-Laser in a chaotic state. The training and testing uses the first 2,000 values of the dataset, divided into three subsets: 1200 (training), 400 (validation), and 400 (test). The first 50 output values of each subset are discarded as an initial washout period. 

The second and third benchmarks are
the Nonlinear AutoRegressive Moving Average model (NARMA), with lags of 10 (NARMA-10) and 30 (NARMA-30) time-steps,
chosen to evaluate the film's ability to manage the nonlinearity--memory trade off~\cite{inubushi2017reservoir}. 
The NARMA task originates from work on training recurrent networks \cite{atiya2000new},
and is now a standard benchmark in reservoir computing. It evaluates a reservoir's ability to model an \textit{N}-th order highly non-linear dynamical system where the system state depends on the driving input $u(t)$ history as well as its own state history $y(t)$. The challenging aspect of the NARMA task is that it contains both non-linearity and long-term dependencies created by the \textit{N}-th order time-lag. 

The $N$th order NARMA equation to be modelled by the reservoir is
\begin{equation}\label{eq:10thNarma}
y(t+1) = \alpha y(t)+\beta y(t)\Bigg(\sum_{i=0}^{N-1}y(t-i)\Bigg) + \gamma u(t-N+1)u(t)+\delta
\end{equation}
For our $N= 10$ and $N=30$ benchmark experiments we use NARMA parameter values of \(\alpha = 0.3\), \(\beta = 0.05\), \(\gamma = 1.5\) and \(\delta = 0.1\) \cite{atiya2000new}.
(The $N= 30$ values are different from the ones first introduced in \cite{Schrauwen2008}.)
An $N$-th ordered NARMA experiment is carried out by predicting the output \(y(t+1)\) given by eqn.\eqref{eq:10thNarma} when supplied with \(u(t)\) from a uniform distribution of interval [0, 0.5].
For each experiment, the NARMA equation is simulated for 6,000 time-steps, with the first 1,000 time-steps discarded as an initial washout period. The remaining 5,000 time-steps are divided into three subsets: 3,000 (training), 1,000 (validation), and 1,000 (test).



\section{Evaluation}\label{sec: Results}
\subsection{Benchmark task results}

\begin{figure}[tp]
\centering
\includegraphics[width=0.95\linewidth,trim=2.5cm 0cm 1cm 0cm,clip]{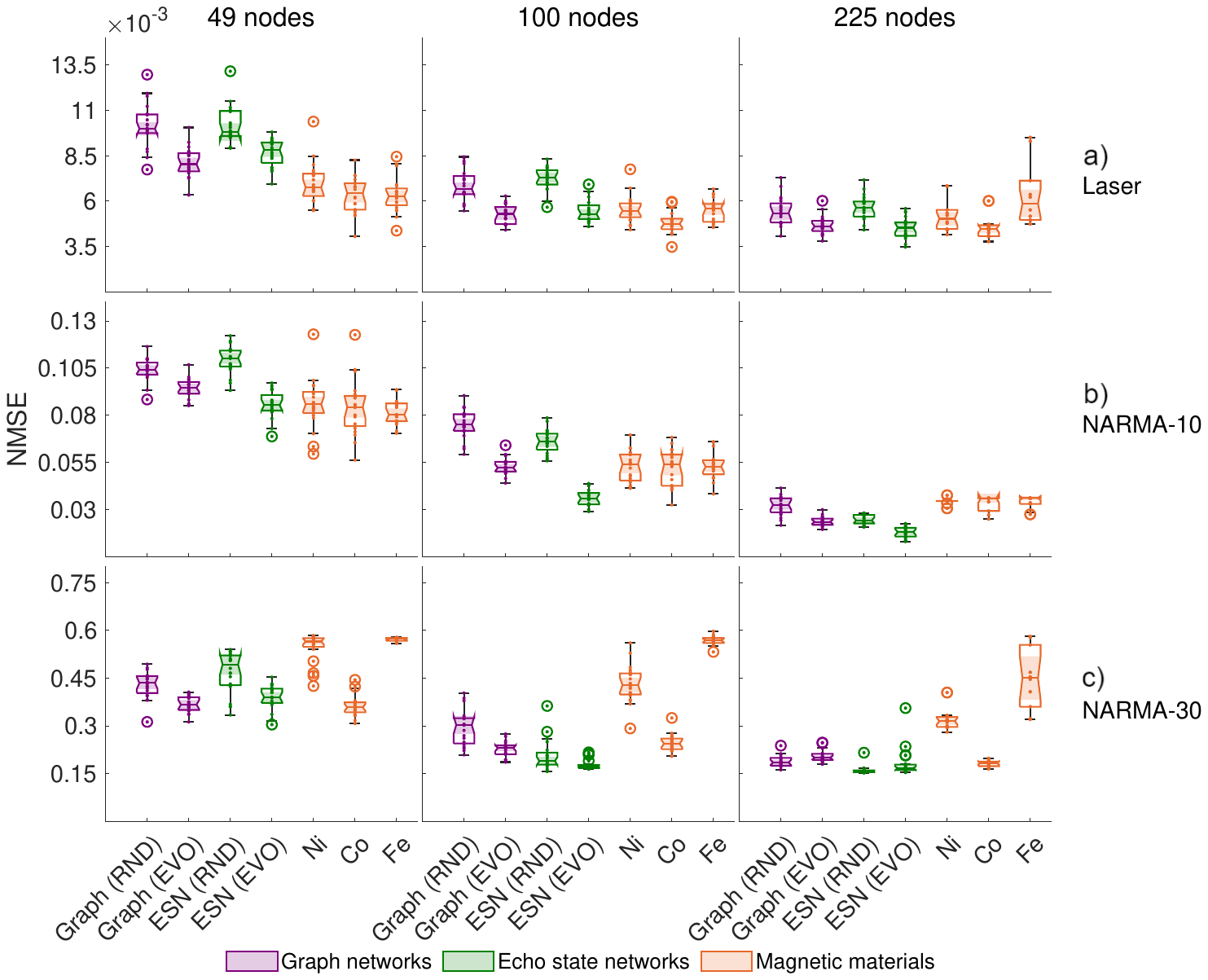}
\caption{Performance of materials and  reservoir networks on benchmark tasks. 
NMSE is used to compare equivalent-sized reservoirs (smaller is better). 
Multiple reservoir sizes are shown in columns, task are shown across rows. Each type of system is represented by colour (graph reservoir$=$purple; echo state reservoir$=$green; material$=$orange). The method used to create the reservoirs is given on the $x$-axis (random or evolved). The boxplots show the best reservoir from each of 20 runs of the search.}
\label{fig: task resullts}
\end{figure}

Fig.~\ref{fig: task resullts} shows the performance of each material at three film sizes. Four results for recurrent neural networks are provided for comparison:  random and evolved searches, each over random and grid connectivity. 
The experimental results show the investigated materials are competitive to  networks, and typically outperform small networks with equivalent reservoir size. 

For the laser task (Fig.~\ref{fig: task resullts}a), all materials significantly outperform random networks at the smallest size of $7\times 7 = 49$ nodes.  
At the intermediate size of $10 \times 10 = 100$ nodes, only Co outperforms evolved networks with a normalised mean square error NMSE of $ 3.5 \times 10^{-3}$, the smallest error found. 
At 100 nodes, Ni and Fe remain statistically similar to evolved networks. For the laser task, even the smallest magnetic reservoirs outperform larger material reservoirs reported in the literature~\cite{larger2012photonic,hou2018prediction}. 
At the largest size ($15 \times 15 = 225$ nodes, right column), only Co outperforms random networks, however, Ni and Fe remain statistically similar. 

For the NARMA-10 task (Fig.~\ref{fig: task resullts}b), all materials outperform random networks at small sizes, except for the evolved ESN. 
At 225 nodes, all materials are statistically similar to random search graph networks but worse than other networks (this is probably an indication of the ineffectiveness of random search compared to evolution). In some cases, materials are better than, or similar to, evolved networks, which have unrestricted access to long-distance connections. The lowest material errors found on this task are an NMSE of 0.056 (Co, 49 nodes), 0.032 (Co, 100) and 0.025 (Co, 225). These are highly similar to, or outperform other, material reservoirs reported in the literature, such as optoelectronic (NMSE $\approx0.168$, 50 nodes~\cite{paquot2012optoelectronic}) and digital reservoirs (NMSE $\approx0.023$, 400 node delay-line ~\cite{appeltant2011information}).

For the first two tasks, the materials  perform comparably.
For the NARMA-30 task (Fig.~\ref{fig: task resullts}c), the difference between materials is clearer.  Co is best able to better match the dynamics of this task. Across all sizes, Co is competitive to random and evolved networks. The lowest error found is a NMSE of 0.165 at 225 nodes. Ni and Fe struggle to compete with other networks at small sizes; nevertheless, as size increases, their NMSE decreases. This suggests that these materials require larger films to exhibit the necessary dynamics to perform the tasks. The reason for this is still unclear and requires further investigation.

\begin{figure}[tp]
\centering
\includegraphics[width=0.95\columnwidth,trim=6.25cm 1cm 4cm 1cm,clip]{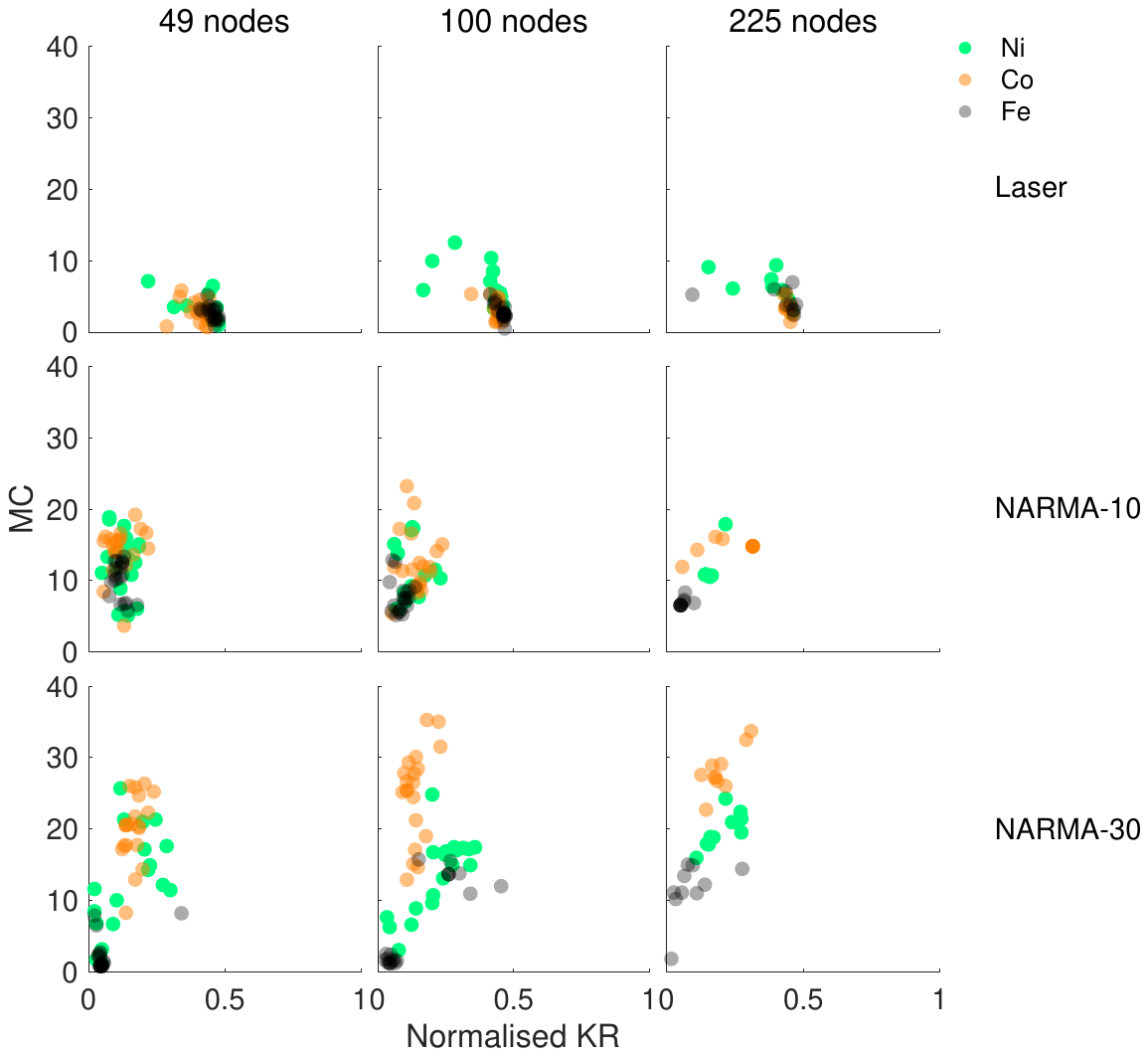}
\caption{Kernel rank (normalised by reservoir size) versus linear memory capacity for all materials, sizes and tasks. Materials are separated by colour: Ni (light green), Co (orange), and Fe (grey). Each column refers to a film size;
each row refers to a task. Material reservoirs shown are the 20 best reservoirs per material per film size per task from the previous experiment. }
\label{fig: all KR and MC}
\end{figure}

\subsection{Nonlinearity and Memory}\label{sec:nonlin}

The results of NARMA-30 task show a strong distinction between the materials, despite their similar performances on other tasks. To understand this further, task-independent measures are used to assess non-linearity and memory. These measures better determine the general underlying dynamics of the system than tasks can achieve alone. They have been used to qualitatively assess the dynamical range of materials for reservoir computing~\cite{dale2019substrate,dale2019role} and to determine a system's total information processing capacity~\cite{dambre2012information}. Here, the non-linear projection and short-term memory are measured, using the kernel rank (KR) and linear memory capacity (MC) of the reservoir. 

Kernel rank is a measure of the reservoir's ability to separate distinct input patterns \cite{legenstein2007edge}. 
It measures a reservoir's ability to produce a rich non-linear representation of the input \(u\) and its history $u(t-1), u(t-2), \ldots$. This is closely linked to the \textit{linear separation property}, measuring how different input signals map onto different reservoir states. As many practical tasks are linearly inseparable, reservoirs typically require some non-linear transformation of the input. KR is a measure of the complexity and diversity of these non-linear operations performed by the reservoir.

Reservoirs in ordered dynamical regimes typically have a low ranking value of KR, and in chaotic regimes, it is high. The maximum value of KR is relative to the number of observable states. In our experiments, KR is normalised to observe the underlying non-linearity of the task without distortion from reservoir size.  

Another important property for reservoir computing is memory, as reservoirs are typically configured to solve temporal problems. A simple measure for reservoir memory is the \textit{linear short-term memory capacity} (MC). This was first outlined in~\cite{jaeger2001short} to quantify the echo state property. For the echo state property to hold, the dynamics of the input driven reservoir must asymptotically wash out any information resulting from initial conditions. This property therefore implies a fading memory exists, characterised by the short-term memory capacity.

A full understanding of a reservoir's memory capacity, however, cannot be encapsulated through a linear memory measure alone, as a reservoir will possess some non-linear memory. Other memory measures proposed in the literature quantify other aspects of memory, such as the quadratic and cross-memory capacities, and total memory of reservoirs using the Fisher Memory Curve~\cite{ganguli2008memory,dambre2012information}. The linear measure is used here as a basic measure of memory capacity. More sophisticated measures are unnecessary to identify the differences in the following tasks.

Fig.~\ref{fig: all KR and MC} shows values of these KR and MC measures for the top 20 material reservoirs discovered in each of the previous experiments. 

The laser task requires very little memory, and is mainly driven by non-linear dynamics. The normalised KR of 0.5 is relatively high when taking into account that many of the magnetic material's observable states are highly correlated, e.g., from the x and y dimension of the spins.

The NARMA-10 task is less non-linear, but requires more memory. Memory capacity clusters around the value of 10, needed for the 10-step time-lag in the system being modelled. Irrespective of size, the same characteristic dynamics have converged during evolution and all materials are able to exhibit the same dynamics.

The NARMA-30 task involves the need for a larger memory capacity still. At the smallest size (49 nodes, with a theoretical upper limit of MC = 49), no material meets the  value necessary to perform well at the task  (MC = 30); 
Co and Ni do manage MCs in the high twenties in some cases, yet Fe struggles to exhibit any memory at all, despite having managed larger values in the other tasks. As size increases, Co and Ni gradually reach \(MC=30\), and this is reflected in their improved performance. The MC of Fe also increases, but at a slower rate proportional to size. This suggests that some property of Fe is less conducive to increasing the MC of the system.

Task performances and KR/MC measure assessment indicate that different trade-offs exist. For example, smaller films generally show better performance than similarly sized digital reservoirs. This suggests properties of small films, such as travelling spin waves reflecting from film edges and interfering (Fig.~\ref{fig: 49 cell input pulse}), may be important for performance. 
Such properties are likely to decrease as film size increases,
indicating that simply scaling up the size of the film may not scale up its computational capacity to the same degree. 
If edge effects are important, the geometry of the film is also likely to have an effect. In our experiments, only square films are used; other shapes can provide greater asymmetry at the boundaries \cite{Dale++:2021-UCNC}. However, depending on the material, larger films may boost desirable dynamical properties such as memory. A large surface area enables signals to persist unperturbed away from rapidly changing input sources. Exploiting geometry, size, and inputs to control these trade-offs are of great interest for future work. 

\subsection{Material parameter value analysis}

As the evolutionary process does not require knowledge of any parameter relationships,  analysis of the evolved parameter values can be performed post-experiment to help understand characteristics of the material and inform future experiments. Evolution is a particularly useful tool when relationships are unknown/uncharacterised/nonlinear for different materials. 

\begin{figure}[tp]
\centering
\subfloat[Laser]{\includegraphics[width=1\columnwidth,trim=2.5cm 6.5cm 1cm 6.5cm,clip]{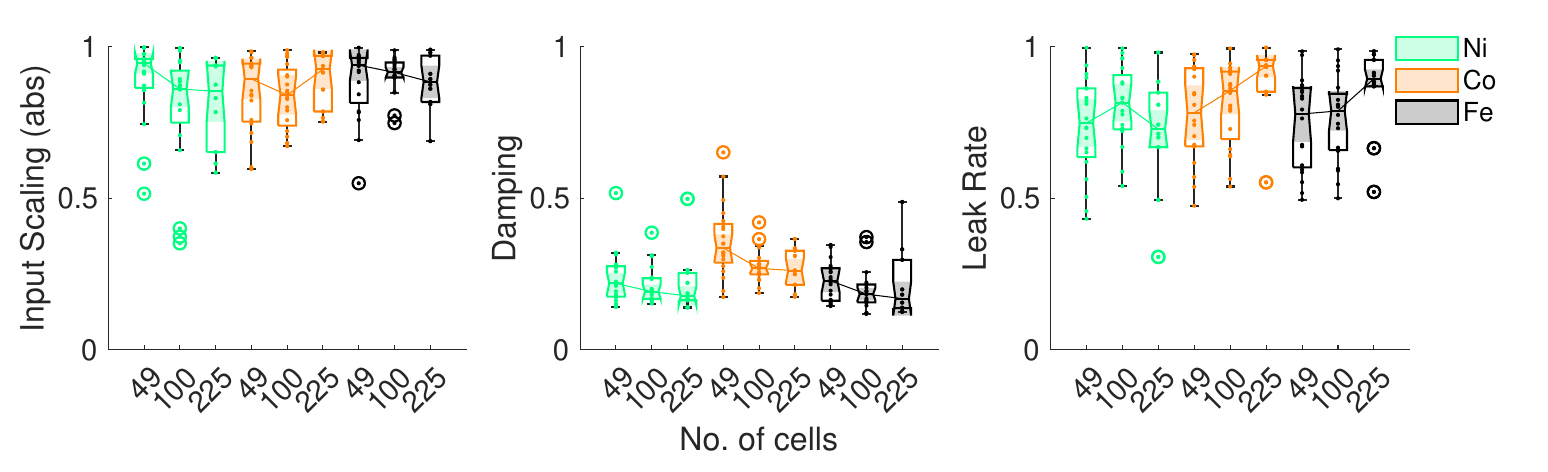}\label{fig: laser param}}

\subfloat[NARMA-10]{\includegraphics[width=1\columnwidth,trim=2.5cm 6.5cm 1cm 6.5cm,clip]{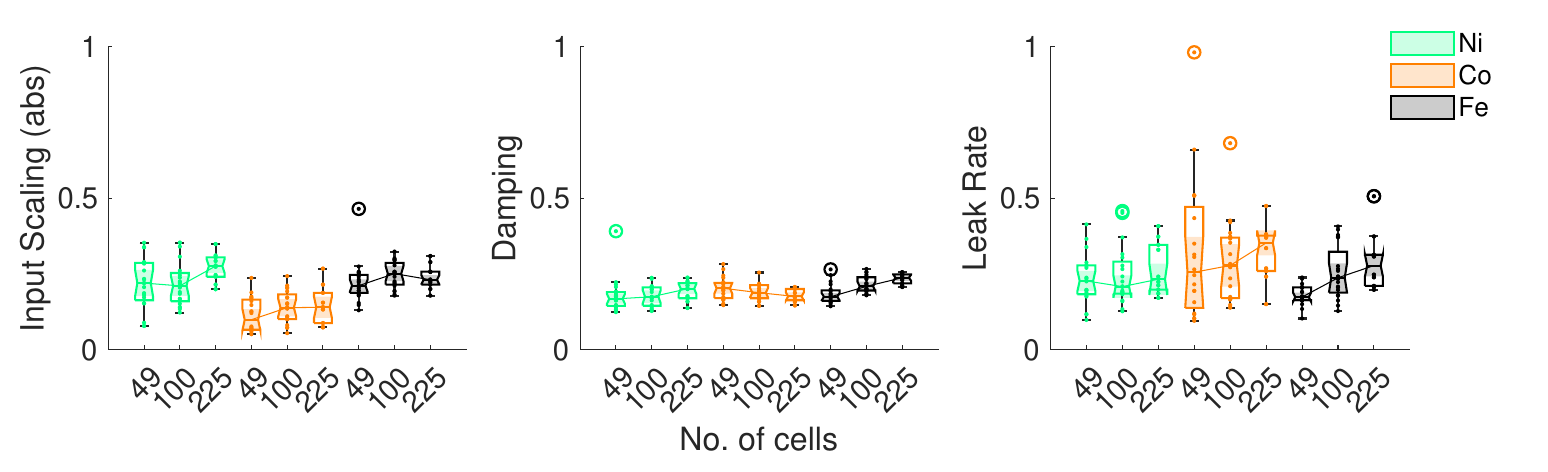}\label{fig: n10 param}}

\subfloat[NARMA-30]{\includegraphics[width=1\columnwidth,trim=2.5cm 6.5cm 1cm 6.5cm,clip]{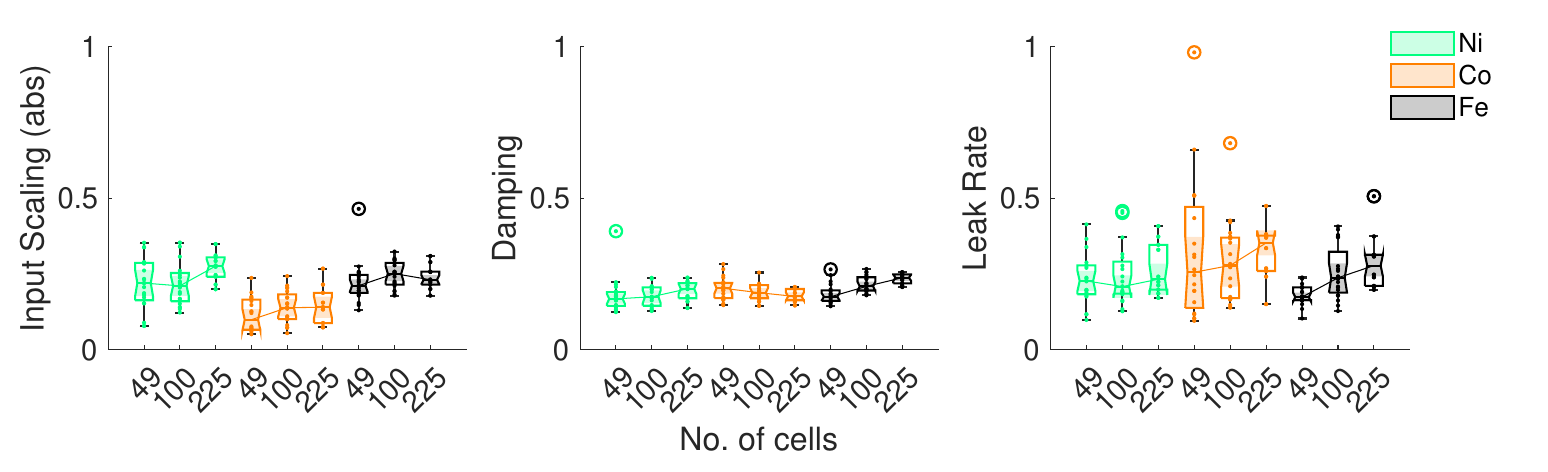}\label{fig: n30 param}}

\caption{Parameters (input scaling $b$, 
damping $\alpha$, leak rate $a$) found by evolution for each material and film size, on each task. 
}
\label{fig: params}
\end{figure}

Each material (Co, Ni, Fe) in our experiments respond differently to parameter values and film size. The best parameter values (of input field scaling $b$, magnetic damping $\alpha$, and leakage rate $a$) discovered in different evolutionary runs for the different tasks and film sizes are shown in Fig.~\ref{fig: params}.

For the laser task  (Fig.~\ref{fig: laser param}), a large input scaling is common among these solutions tending towards 1 (positive and negative) with absolute value averages of 0.86 (Ni), 0.84 (Co), and 0.91 (Fe) for 100 nodes. 
Damping tends to be low on average: 0.19  (Ni), 0.26 (Co), and 0.18 (Fe). 
The leak rate tends to be high: 0.82 (Ni), 0.85 (Co), and 0.79 (Fe). As size increases, input scaling and damping tend to decrease across all materials whilst leak rate tends to increase.  

For the NARMA-10 task (Fig.~\ref{fig: n10 param}), a low input scaling is common with averages of 0.20 (Ni), 0.13 (Co), and 0.25 (Fe) at 100 nodes. 
Damping tends to be low on average: 0.17 (Ni), 0.18 (Co), and 0.20 (Fe). 
The leak rate also tends to be low: 0.20 (Ni), 0.27 (Co), and 0.23 (Fe). 
As size increases, there are different movements in parameters, depending on the material. Leak rate typically increases, input scaling increases slightly, and damping increases for Fe and Ni, but not Co. 

For the NARMA-30 task (Fig.~\ref{fig: n30 param}), a low input scaling is common again with averages of 0.15 (Ni), 0.05 (Co), and 0.37 (Fe) at 100 nodes. 
Damping tends to be low on average: 0.10 (Ni), 0.12 (Co), and 0.43 (Fe). 
The leak rate tends to be intermediate: 0.57 (Ni), 0.46 (Co), and 0.1 (Fe). 
The damping and input scaling ranges are generally lower than the NARMA-10 task, suggesting lower values promote greater memory. There are also significant changes in values as size increases for the Ni and Fe materials. This reflects the challenge to find suitable parameters that provide low errors for these materials at smaller sizes. The low variance as size increases also shows the convergence to suitable parameters becomes easier.

Overall, taking into account tasks errors, task-independent measures, and average parameter values, we can infer a general rule that a high input scaling creates a higher KR and lowers memory, and vice versa. As damping decreases, memory increases when accompanied by a low input scaling. And, a lower leak rate can benefit memory. This is typical for all materials at different sizes, however, averages increase or decrease depending on the material and task. The Fe parameters do not fit the trend for the NARMA-30 task because evolution struggles to find parameters that produce the necessary dynamics.

\section{Possible Paths to Physical Realisation} \label{sec:paths} 

\begin{figure}[tp]
\centering
\includegraphics[width=0.95\linewidth,trim=2.5cm 0.75cm 3.25cm 0cm,clip]{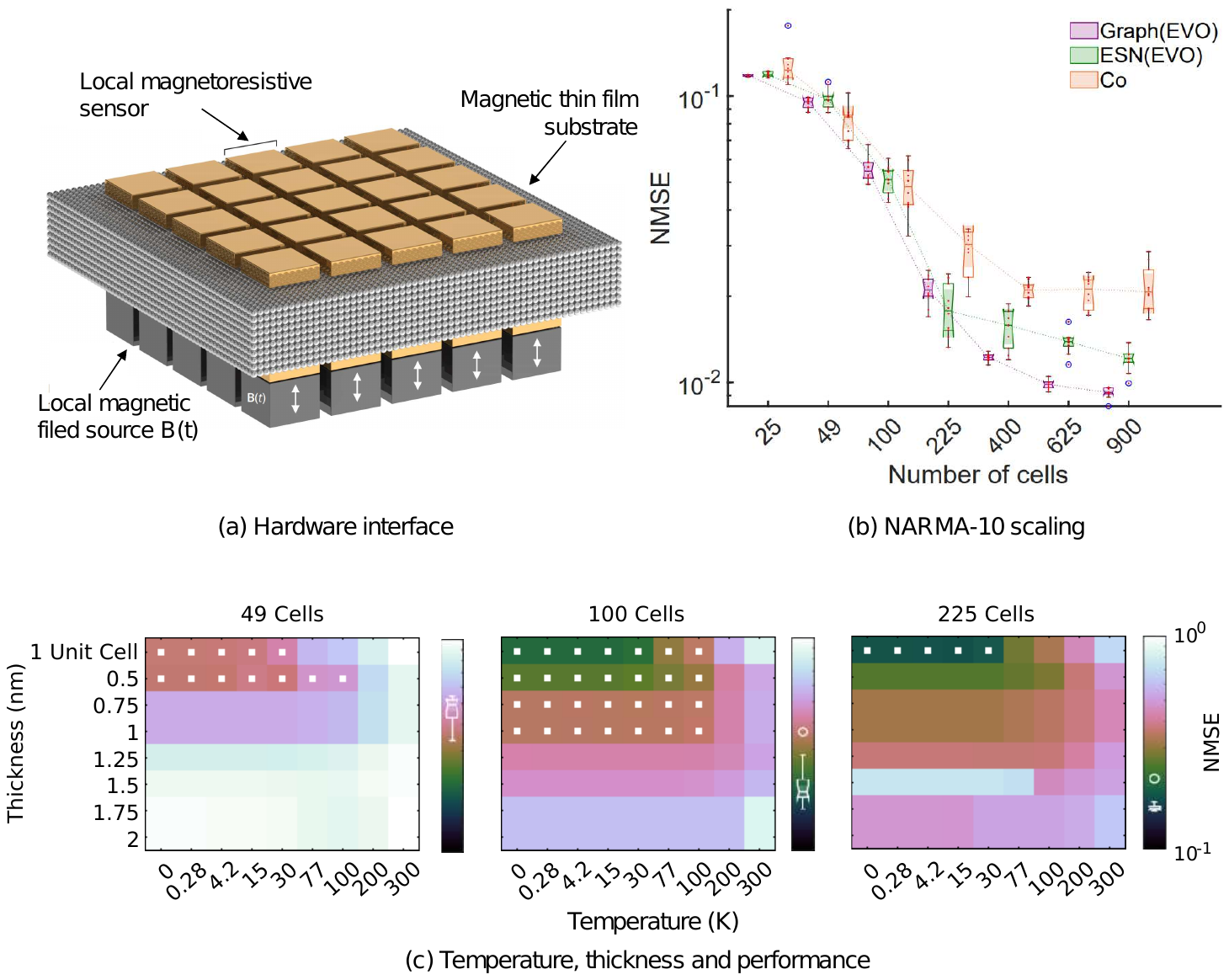}
\caption{
({a}) Possible hardware interface to realise a thin-film reservoir computing device. Input signals are induced by local magnetic field sources $B(t)$ and the reservoir states of the film are read via local magnetoresistive sensors. 
({b}) Performance of Co material on the NARMA-10 task as number of macro-cells increases. 
({c}) Grid sweep of film temperatures (K) and film thickness (nm). The best previously evolved Co configuration (evolved at 0\,K and one unit cell $\approx$ 0.25\,nm thickness) is evaluated at different temperatures and thicknesses.
The NMSEs are for the NARMA-30 task. Temperature ranges from 0\,K (original experiments) to room temperature (300\,K).
Thicknesses range from one unit cell,  the thinnest film that can be simulated, to 2\,nm.
A white square indicates that performance of the film is within the ESN range.
The box plot in the colour bar show the performances of the best random ESN in each of the 20 runs. 
} 
\label{fig: scaling and temperature}
\end{figure}

The simulated platform is in principle realisable in physical hardware, but advances in technology may be required to implement the films and system interface at the required spatial scales and frequencies. 
Fig.~\ref{fig: scaling and temperature}a shows a possible 5$\times$5 input-output interface. The device consists of a nanoscale thin-film, encapsulated by point contacts (yellow) that measure the local tunnelling magnetoresistance (i.e. the reservoir state) in different regions of the film. Underlying magnetic field sources (grey) provide locally controllable magnetic field inputs $B(t)$ via electrical currents to each region of the device. Each source is independent and has an additive effect on the local magnetic state of the film. The full details of the physical device are not covered in this theoretical work and remain future research.

To explore the viability of the system, further key aspects are covered here: 
film size, temperature, and film thickness. 

\subsection{Film size} 

With any new reservoir system, an ability to scale hardware components and reduce error is desired. In our experiments, each material exhibits improvement as film size increases (Fig.~\ref{fig: task resullts}), despite indications that edge effects might be playing a role (section~\ref{sec:nonlin}). The greatest improvements relate to the difficulty of the task, where distinct trade-offs in non-linearity and memory are required. The most significant differences between material and size are shown for the NARMA tasks, where memory is a strong indicator of performance. 

To assess scaling potential, we performed additional evolutionary searches with the Co material for larger systems. In order to compare material scaling with digital reservoir scaling, we also evolved equivalent-sized networks. Fig.~\ref{fig: scaling and temperature}b shows performance on the NARMA-10 task as film and reservoir size increases. Scaling runs from $5\times 5 = 25$ material macro-cells corresponding to  25\,nm $\times$ 25\,nm films, up to $30\times 30 = 900$ macro-cells corresponding to  150\,nm $\times$ 150\,nm films.
The results show that up to 400 macro-cells/network nodes there is a significant reduction in the average error as size increases, and the material remains comparable to simulated networks. After this, the material's median error is no longer significantly decreasing (although lower errors continue to be found in the best runs), whilst the simulated networks continue to improve with size. 
This could indicate that larger films with lower errors are more challenging to discover, or that beneficial properties of small films are lost, such as interaction of reflections from edges \cite{Dale++:2021-UCNC}.

This indicates that simple size scaling may not  be a route to make more powerful RCs for magnetic thin films: bigger is not necessarily better. Instead, scaling may need to be accomplished by combining results from multiple magnetic films with different properties, or by using different shapes of magnetic thin films. While regular shapes are useful in the context of simulation, the use of multiple smaller films with potential irregularities to their shape may simplify some of the manufacturing challenges. However, this is beyond the scope of this initial work.

\subsection{Temperature, and film thickness} 

At the nanoscale, thermal noise is a limiting factor. Maintaining performance close to room temperature is desirable for practical implementations. Stability and reproducibility can be adversely affected by thermal noise. In our experiments, temperature is set to absolute zero kelvin to observe pure magnetic behaviour without thermal effects. Methods to control and reduce thermal fluctuations have been proposed using spin transfer torque to modify thermal activation rates~\cite{demidov2012magnetic}. This suggests different paths towards room temperature computing without cooling are plausible.

To demonstrate the effect of temperature on our films, we performed additional experiments. Fig.~\ref{fig: scaling and temperature}c shows reservoir performance at various temperatures on the NARMA-30 task, using the optimal parameter values found by evolution for the 0\,K case. The temperature range includes: millikelvin ($0.28$\,K), liquid helium ($4.2$\,K), liquid nitrogen ($77$\,K), and room temperature ($300$\,K). The top-left shows the original experimental setup (temperature = 0\,K and film thickness = one unit cell $\approx0.25$nm, see table~\ref{tab:material param}) for an evolved Co reservoir. As temperature increases along the $x$-axis, thermal noise dominates and degrades performance. A similar pattern is present across all film sizes, tasks and materials (see Appendix~\ref{app: additional temp and scaling results}). 

Film thickness is also investigated to see whether thickness can compensate for a rise in temperature. On the $y$-axis of Fig.~\ref{fig: scaling and temperature}c, film thickness varies from 0.25--2\,nm. In general, performance is maintained with thicknesses up to $0.5$\,nm and temperatures up to 30--77\,K. Between 0.5--1\,nm, the change in error slows as temperature rises (30 to 200\,K), however errors are higher than for thinner films. Beyond $1$\,nm, thicker films tend to degrade performance, but this varies depending on material and film size (see Appendix~\ref{app: additional temp and scaling results}). The results show that films with sub-nanometer thickness at temperatures up to 100\,K work best, outperforming or matching equivalent-sized random reservoir networks.
Further work is needed to evolve parameter values specific to different temperatures and film thicknesses, to see how much performance can be improved thereby.

Using a thinner film  to compensate for thermal noise carries other drawbacks, however.
A 0.5\,nm film is
at the upper end of thickness
giving an acceptable computational performance within a reasonable temperature range; it 
would have significant fabrication challenges, especially with regard to the effect of impurities. 
While impurities might produce benefits by introducing additional non-linear behaviours, they are beyond the scope of this initial exploration. 
Other ferromagnetic materials can also be characterised with the same techniques, and these materials may present better characteristics, or other desirable properties (such as 
being easier to work with). Hence, a key challenge for fabrication would be to determine a suitable material to use for the thin film magnetic layer.

\section{Discussion}

The chosen benchmarks are ones commonly used to evaluate reservoir systems, but  provide only a snapshot of the investigated systems' capabilities. 
The tasks vary in difficulty, as seen by the KR and MC in Fig.~\ref{fig: all KR and MC}, and require only a few inputs. Although we show competitive performance in simulation to digital reservoirs of comparable size, the matching of system dynamics to task dynamics is critical for good task performance; as demonstrated for the NARMA-30 task. %

This matching process highlights a fundamental challenge when exploring and proposing any new unconventional computing system: that determining or understanding the general computing capabilities of physical reservoirs using benchmark tasks alone is fraught with potential problems.

Physical systems are inherently more constrained in terms of dynamics due to their underlying physics and structure. Even basic constraints such as the number of physical inputs-outputs and internal time-scales can affect the choice of tasks. 


Computing problems and tasks vary in requirements and difficulty. Nonlinearity, memory, and sometimes ``edge of chaos'' dynamics (criticality) are considered essential for reservoir computing, yet tasks fit on a spectrum of required dynamics. Linear systems can perform some tasks surprisingly well~\cite{Gallicchio2019} and Extreme Learning Machines (ELM)~\cite{butcher2013reservoir,ortin2015unified} (random neural networks without memory) can perform many computing tasks without the echo state property. The ``edge of chaos'' -- or rather the ``edge of stability'' as described in~\cite{carroll2020reservoir} -- is not necessary in all scenarios either. In typical RC benchmarks, the nonlinearity and memory required for different tasks varies considerably as a trade-off exists between the two~\cite{inubushi2017reservoir,dambre2012information}. 
On another axis we have task difficulty. What is an appropriate/acceptable level of task difficulty to determine how useful a system is? In unconventional computing, solving the XOR problem is common to show the presence of nonlinearity, as a nonlinear decision boundary is essential to solve it. However, the task is simple and can be solved with only two sigmoidal neurons, thus providing little information about the complexity of tasks it might eventually perform, unless the original objective is to build only unconventional XOR gates.

Another challenge is identifying what parts of the system are computing and how much each part is contributing to the final solution. This is critical when proposing any low-power alternative or real-time system that intends to perform the bulk of the computing, i.e., if considerable (digital) preprocessing is required, this additional step comes at a cost, or can lead to bottlenecks; similar to the memory retrieval (von Neumann) bottleneck in conventional computers. For example, experiments demonstrating speech recognition with reservoir computers show good performance, yet a considerable bulk of the ``heavy lifting'' is done at the preprocessing and feature reduction/extraction stage~\cite{araujo2020role}.

In practice, physical reservoir systems (single device architectures and more exotic reservoirs at least) may perform well only in specific task domains and functionality outside these domains is likely to be suboptimal. Yet, knowing what is a suitable task domain from the outset is difficult. To understand what is an appropriate task for any unconventional computing system, we need to know the range of dynamics the system can exhibit. Task-independent measures such as the kernel rank, capacity to reconstruct a function, entropy and mutual information, and linear and nonlinear memory capacities ~\cite{legenstein2007edge,boedecker2012information,dambre2012information,torda2018evaluation} are more informative about the system's potential than task-specific benchmarks. 

The CHAracterisation of Reservoir Computers (CHARC) framework \cite{dale2019substrate} is a tool to integrate such measures to experimentally determine the system's dynamical range and capabilities. With this, one can compare reservoir systems and identify suitable tasks. At a basic level, the framework maps the dynamical freedom of the system when applying design choices (e.g. input encoding, physical constraints, computational model, etc.) highlighting what the system can and cannot exhibit. From this, an in-depth understanding of the parameter space and space of all excitable dynamics is possible, which enables one to choose (or even predict) suitable applications. Once these properties are better understood, and can be evaluated and compared easily, it may be possible to guide and iteratively improve system design~\cite{dale2020design}, from material properties to the computing model. 

\section{Conclusion}

Our spintronic-based system provides a novel substrate for machine learning with analogue hardware. By combining two frameworks, evolution \textit{in materio} and reservoir computing, novel magnetic computing devices are demonstrated.

Without the need for discrete neural components, physical reservoirs are possible with smaller footprints than other neuromorphic devices, e.g., memristors, spin torque oscillators, photonics~\cite{du2017reservoir,torrejon2017neuromorphic,romera2018vowel}. The evolved devices operate at frequencies of 100 GHz and require no special preprocessing to emulate network structures~\cite{appeltant2011information,torrejon2017neuromorphic}. The basic materials used are inexpensive and feature a large dynamical range that can be reconfigured externally to solve different machine learning tasks. 

Some engineering challenges still remain, including fabricating the high-speed interface to exploit the natural timescales of the materials. These timescales are tunable to some extent and other materials will feature slower or faster internal dynamics, e.g. antiferromagnetic materials (THz)~\cite{kurenkov2020neuromorphic}. There is also some possibility of operating at slower speeds (1--5\,GHz) and even quasi-statically. Research into THz signal modulation and sensors~\cite{jin2020magnetic,wang2019review,zhang2020ultrafast} will help future construction of the platform. 
There are also questions on how  the practical difficulties of working with such thin small magnetic films can be realistically overcome, 
and which other magnetic materials might be exploited.
The materials simulated here are `ideal' uniform materials; inhomogeneities and imperfections are not included.  Our prior experience with extremely inhomogeneous materials \cite{dale2016reservoir,dale2017reservoir,dale2019substrate}, however, does not lead us to expect major changes in the conclusions: all that is needed for this style of RC is a system with suitably rich non-linear dynamics.

Three standard benchmarks are used to evaluate and compare each material to simulated networks. Material performances are shown to be highly competitive to optimised recurrent neural networks across different sizes. 
However, to achieve a full comprehensive characterisation of the materials, and evaluate other features such as geometry, frameworks like CHARC~\mbox{\cite{dale2019substrate,dale2019role}} which explore the full dynamical range of the reservoir system will be the focus for future work.

It is possible to extend our approach to other magnetic materials which may have more desirable properties, including more complex materials such as alloys, oxides, skyrmion fabrics, spin-torque oscillator arrays \cite{Dale2023-UCNC}, and anti\-ferro\-magnetic reservoirs~\cite{kurenkov2020neuromorphic}. 
The natural dynamics and nanoscale size of these various magnetic substrates presents a new path towards fast efficient computing platforms enabling new innovations in smart technologies.

%% file: supplementary.tex
\section{Optimised Integration Time-step} \label{app: timestep}
\begin{figure}[tp]
\centering
\includegraphics[width=0.9\columnwidth,trim=2cm 6cm 2cm 6cm,clip]{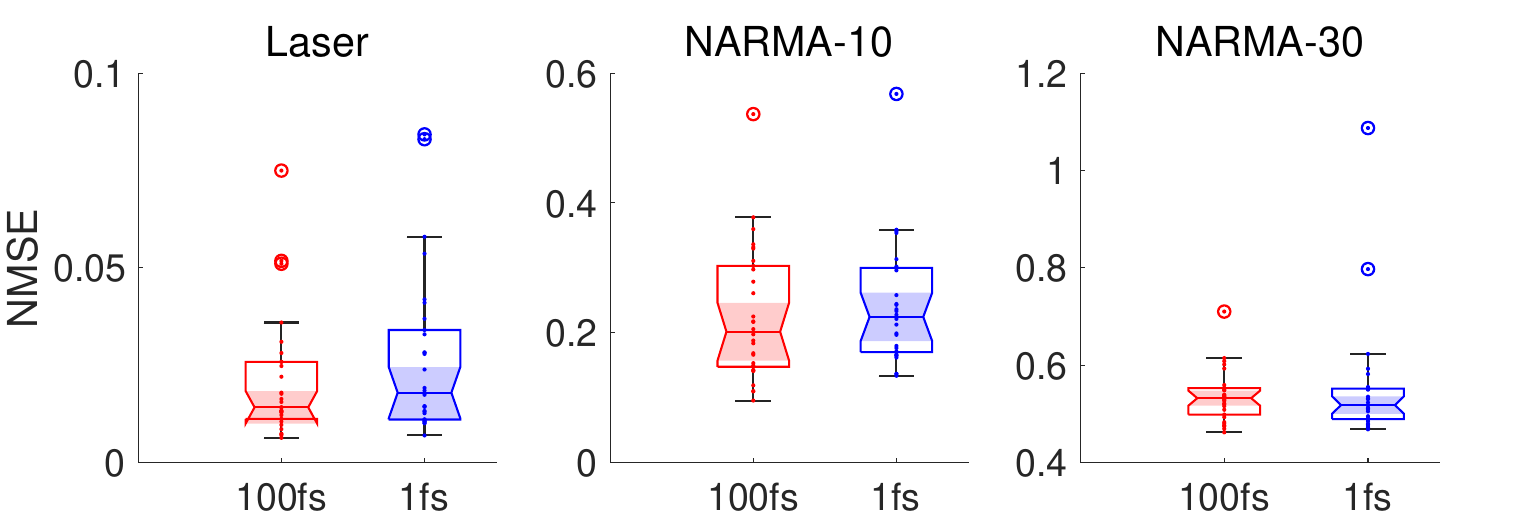}
\caption{Comparing integrator time-steps across three tasks. 100\,fs provides a less accurate model compared to 1\,fs but dramatically reduces run-time. The Co material used in this experiment. Each boxplot shows a total of 30 random configurations for each task compared at both time-steps.}
\label{fig: ts comparison}
\end{figure}

To reduce computational time simulating thin-films a large integrator time-step was used. A characterisation of the how the integrator time step affects task performance is given in Fig.~\ref{fig: ts comparison}. The results show that the chosen 100\,fs time-step is statistically similar to the 1\,fs time-step, representing a reasonably accurate model of the driven dynamics.

To test whether the medians of both time-steps were significantly different, the non-parametric two-sided Wilcoxon rank sum test was used. This tests the null hypothesis that both samples are from the same distribution with equal medians. A rejection of the null hypothesis at the 95\% significance level is indicated by a \textit{p}-value $>0.05$. The \textit{p}-values for each task are: $p=0.23$ (laser), $p=0.42$ (NARMA-10), and $p=0.57$ (NARMA-30). This indicates that performance is not significantly affected by the change in time-step, however, computational time is reduced dramatically from hours to minutes.

\section{Temperature Effect and Film Thickness} \label{app: additional temp and scaling results}
To build practical computing systems it is desirable for the materials to function close to room temperature. In addition, thicker films put less strain on the fabrication process. In our main experiments, each material film was evolved at 0\,K to evaluate performance without thermal fluctuations. Here, we show how temperature affects performance at all film sizes (number of cells) and across each task.

\begin{figure}[tp]
\centering
\includegraphics[width=0.95\columnwidth]{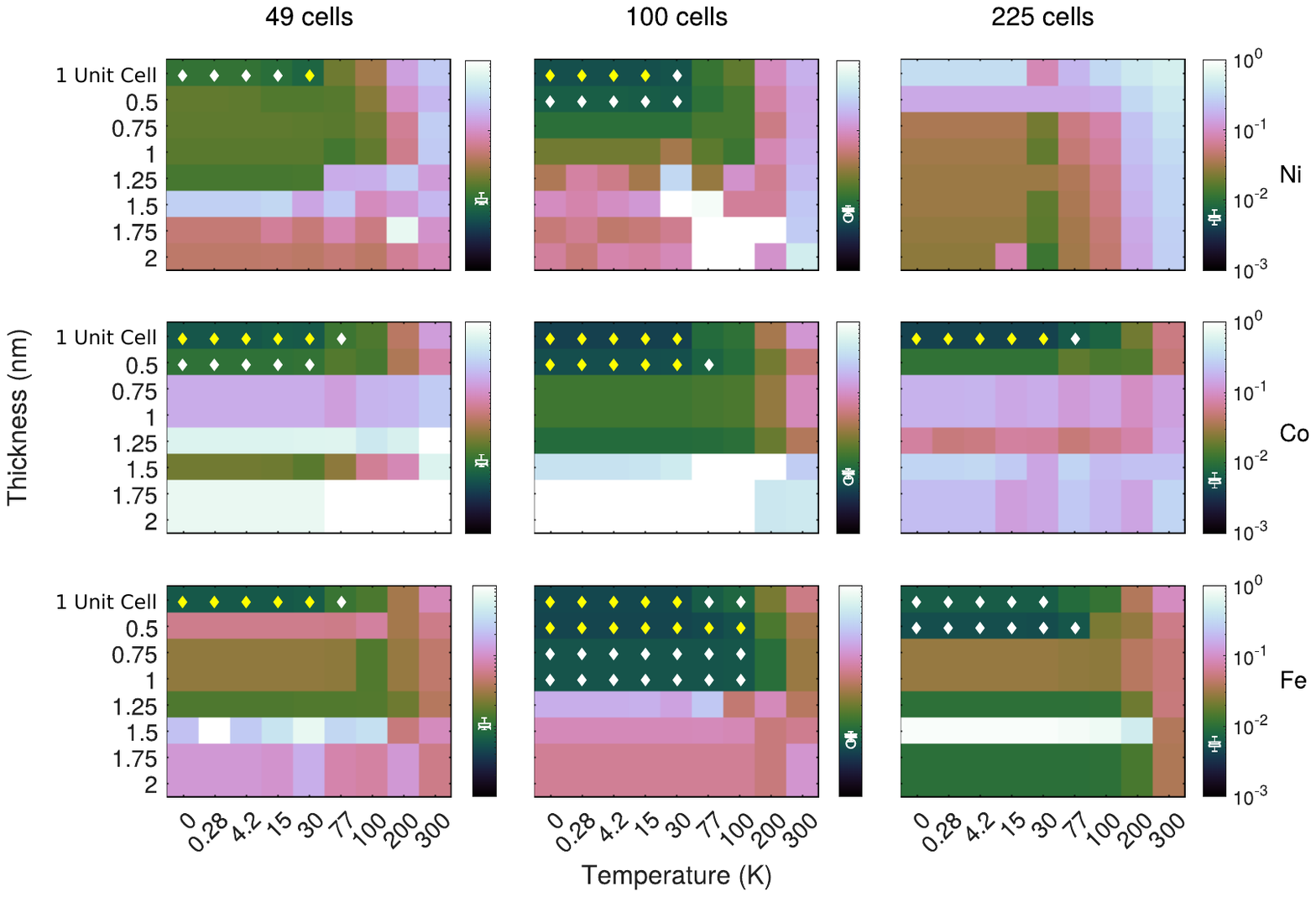} 
\caption{\label{fig:laser temp}Temperature and film thickness sweeps for the Laser task. Normalised mean square error (colour; darker is better) of an evolved configuration for Ni (top), Co (middle), Fe (bottom). 
``1 Unit Cell" is the thinnest film that can be simulated; see table~\ref{tab:material param} for the thickness this corresponds to for each of the materials.
A white diamond indicates that performance of the film is within the ESN range. A yellow diamond indicates that performance is better than the best ESN. 
The box plot in the colour bar show the performances of the best random ESN in each of the 20 runs. 
}
\end{figure}

For the laser task (Fig.~\ref{fig:laser temp}), performance is stable and competitive -- to random ESNs of equivalent node size -- at higher temperatures typically up to 100K, depending on the material and number of cells. The most stable material and film size is Fe at 100 cells. In this configuration, only a small change in performance is present as thickness is increased up to 1nm.  

\begin{figure}[tp]
\centering
\includegraphics[width=0.95\columnwidth]{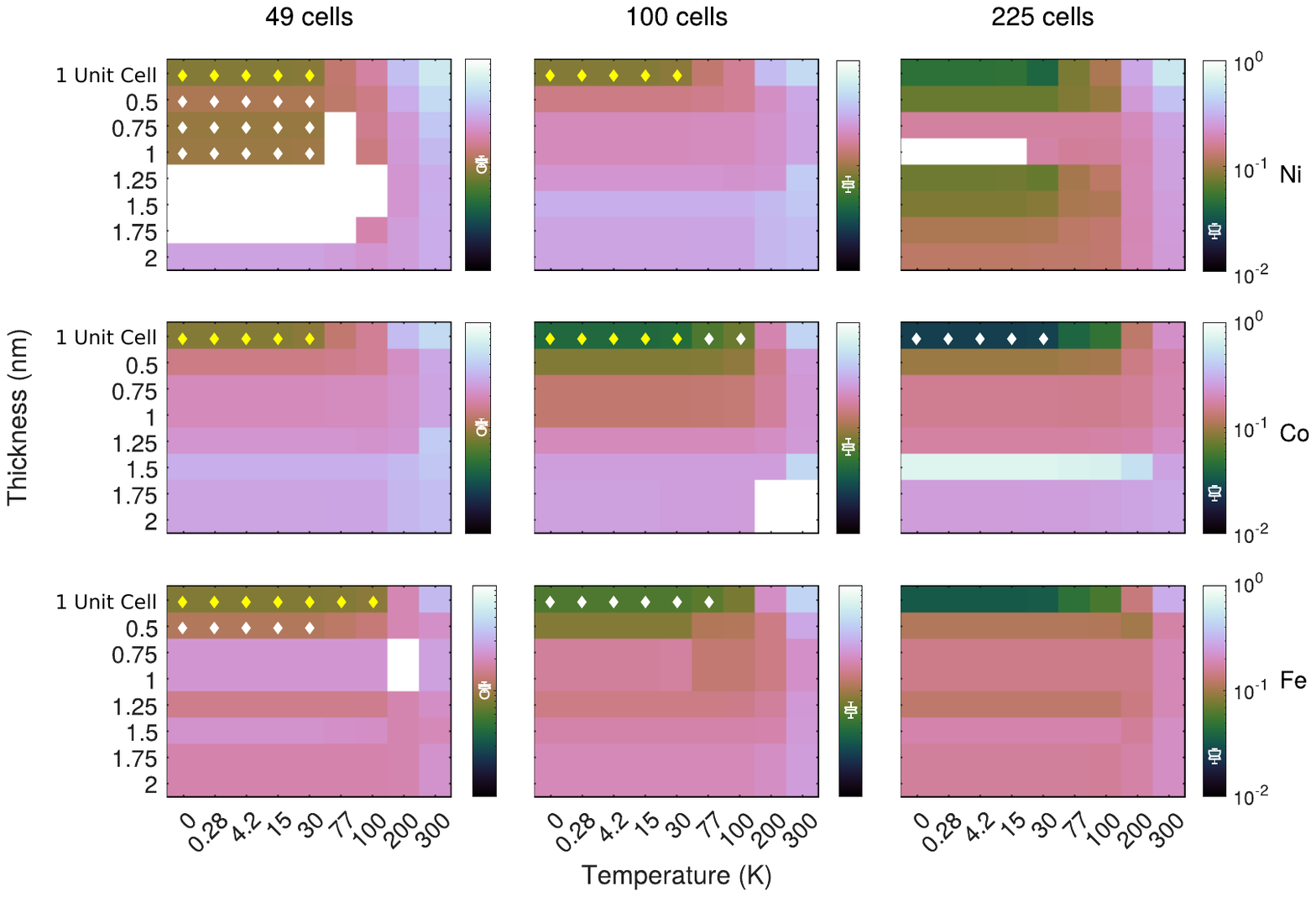}
\caption{\label{fig:n10 temp}Temperature and film thickness sweeps for NARMA-10 task.
See figure~\ref{fig:laser temp} caption for key.
}
\end{figure}

\begin{figure}[tp]
\centering
\includegraphics[width=0.95\columnwidth]{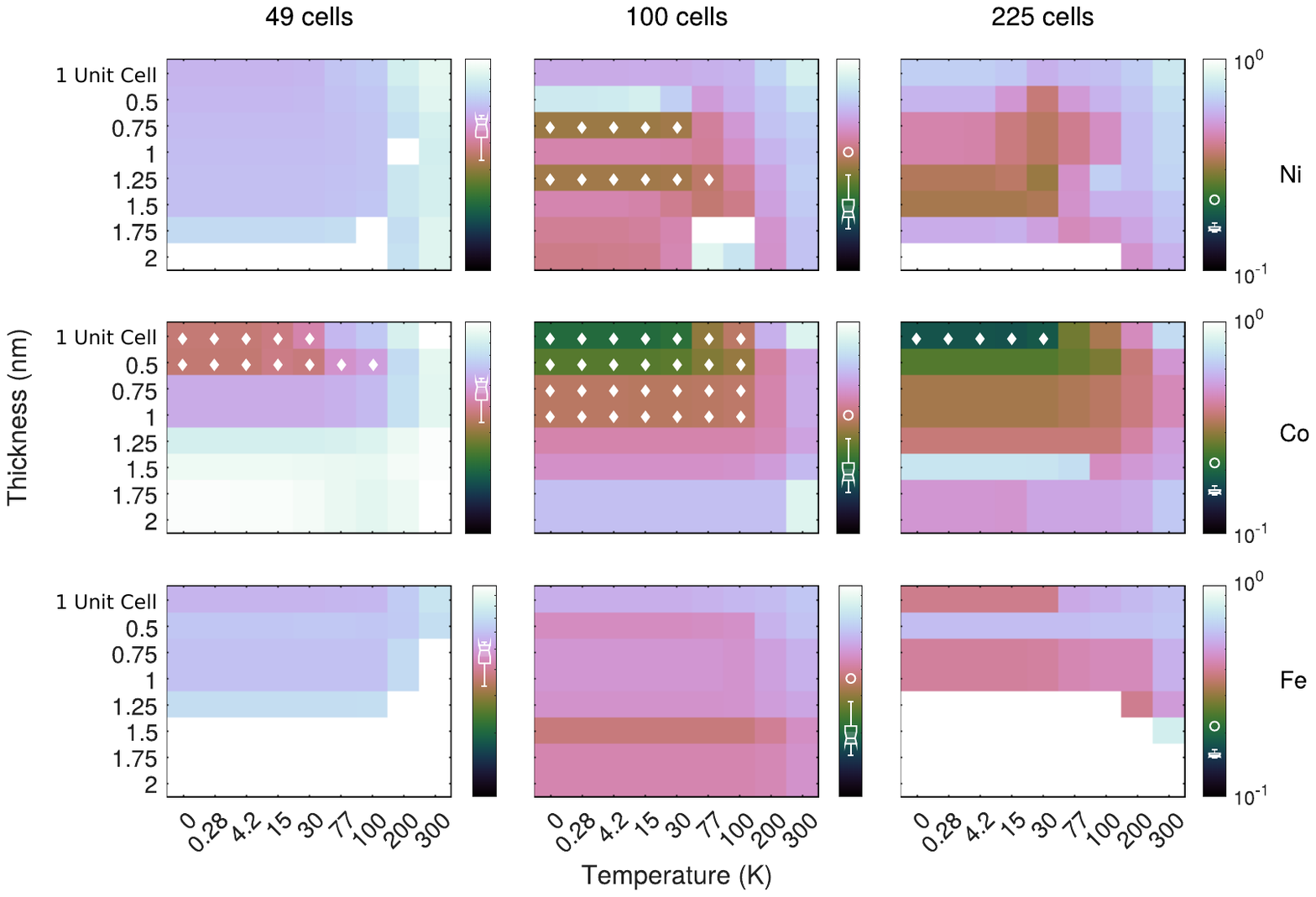}
\caption{\label{fig:n30 temp}Temperature and film thickness sweeps for NARMA-30 task. 
See figure~\ref{fig:laser temp} caption for key.
}
\end{figure}

For the NARMA-10 task (Fig.~\ref{fig:n10 temp}) performance is again stable, in some cases up to 100K, e.g., Co with 100 cells. As temperature increases, performance tends to drop off slightly faster than the laser task. This could be due to degradation in memory quality as thermal noise increases. In general, the results suggest the Co material responds better to increased temperatures. However, thicker films tend to be more detrimental to performance. The same trends are seen for the NARMA-30 task  (Fig.~\ref{fig:n30 temp}).

